\documentclass[preprint2]{emulateapj}

\usepackage{subfigure}
\usepackage{tabularx}
\usepackage{amsmath}
\usepackage{amssymb}

\shorttitle{GES in inner-central halo of MW}
\shortauthors{Kunder et al.}

\begin{document}

\title{The Galactic Bulge exploration VI.: Gaia Enceladus/Sausage RR Lyrae stars in the inner-central stellar halo of the Milky Way}

\author{
Andrea Kunder\altaffilmark{1},
Zdenek Prudil\altaffilmark{2},
Antonela Monachesi\altaffilmark{3},
Samuel J. Morris\altaffilmark{1},
Kathryn Devine\altaffilmark{4},
Joanne Hughes\altaffilmark{5},
Kevin Covey\altaffilmark{6},
R. Michael Rich\altaffilmark{7},
Elisa A. Tau\altaffilmark{8}
}

\altaffiltext{1}{Saint Martin's University, 5000 Abbey Way SE, Lacey, WA, 98503, USA}
\altaffiltext{2}{European Southern Observatory, Karl-Schwarzschild-Strasse 2, 85748 Garching bei M\"{u}nchen, Germany}
\altaffiltext{3}{Departamento de Astronomía, Universidad de La Serena, Av. Ra\'{u}l Bitr\'{a}n 1305, La Serena, Chile}
\altaffiltext{4}{The College of Idaho, 2112 Cleveland Blvd Caldwell, ID, 83605, USA}
\altaffiltext{5}{Physics Department Seattle University, 901 12th Ave., Seattle, WA 98122, USA}
\altaffiltext{6}{Department of Physics \& Astronomy, Western Washington University, MS-9164, 516 High St., Bellingham, WA, 98225}
\altaffiltext{7}{Department of Physics and Astronomy, UCLA, 430 Portola Plaza, Box 951547, Los Angeles, CA 90095-1547, USA}
\altaffiltext{8}{Departamento de Astronomía, Universidad de La Serena, Av. Ra\'{u}l Bitr\'{a}n 1305, La Serena, Chile}

\begin{abstract}
We present a view of the stellar halo in the inner-central regions of the Milky Way ($R\rm  \lesssim 10~kpc$)  mapped by RR Lyrae stars.   
The combined BRAVA-RR/APOGEE RR Lyrae catalog is used to obtain a sample of 281 RR Lyrae stars located in the bulge region of the Galaxy, but with orbits indicating they belong to the inner-central halo.  
The RR Lyrae stars in the halo are more metal-poor than the bulge RR Lyrae stars and have pulsation properties more consistent with an accreted population.  
We use the Milky Way-like zoom-in cosmological simulation Auriga to compare the properties of the RR Lyrae stars to those expected from the “Gaia-Enceladus-Sausage” (GES) merger.  The integrals of motions and eccentricities of the RR Lyrae stars are consistent with a small fraction of 6-9 $\pm$ 2 \% of the inner-central halo RR Lyrae population having originated from GES.  
This fraction, lower than what is seen in the solar neighborhood, is consistent with trends seen in the Auriga simulation, where a GES-like merger would have a decreasing fraction of GES stars at small Galactocentric radii compared to other accreted populations.  
Very few of the Auriga inner Galaxy GES-18 particles have properties consistent with belonging to a bulge population with ($z_{max} <$ 1.1~kpc), indicating that no (or very few) RR Lyrae stars with bulge orbits should have originated from GES.
\end{abstract}


\section{Introduction} 
Galaxies are embedded within halos -- extended, roughly spherical mass distributions dominated by dark matter, but also consisting of tenuous gas and a sparse stellar component \citep[$e.g.,$ the Milky Way halo consists of $<$1\% of the total stellar mass of the Galaxy,][]{deason19, girelli20}.  
The formation of halos in galaxies arise naturally in a $\Lambda$ cold dark matter ($\Lambda$CDM) paradigm through hierarchical mass assembly \citep[$e.g.$,][]{bullock05, cooper10}.  
Accretion events are the building blocks of halos, and these past mergers map the mass assembly history of the Galaxy \citep[$e.g.$,][]{belokurov18, kruijssen20}.  

Stellar debris from past accretion events tend to be most easily seen in regions of the Galaxy with little stellar contamination from other populations; stellar associations can be seen without confusion by the general field stars, and also where phase-mixing and dissolution of sub-structure is not as rapid.  As such, the stellar halo of the Milky Way Galaxy at small Galactocentric radii ($R \rm \lesssim 6~kpc$) is typically avoided when probing the properties of the Galactic halo.  However, according to simulations, the stellar halo will be more prominent in highly dense areas, so the halo extending from the dense Galaxy center would be important to probe \citep[$e.g.$,][]{monachesi19}.  
While it is not always clear how deep into the Milky Way's potential 
accreted stars will penetrate as compared to stars formed $in~situ$ at late times \citep{elbadry18}, 
accretion events will eventually in-spiral toward the Galactic center driven by dynamical friction \citep[][]{amorisco17}. 
Thus, the inner region of the Galaxy is a strategic location to uncover and study remnants for mergers.  

One of the more significant mergers in the Milky Way, which has been shown to dominate the stellar halo in the inner (6~$ <~R~<$~25~kpc) regions, is the Gaia-Enceladus/Sausage (GES) merger \citep{helmi18, belokurov18}.  
GES happened 8-11~Gyr ago, and was a massive merger ($\rm M_{*} \sim 10^9 M_{\odot}$) bringing in stars with mean metallicities of $\rm [Fe/H] \sim -$1.3 with $\rm [Mg/Fe]$ and $\rm [Al/Fe]$ lower than the $in~situ$ Milky Way stars at that given metallicity \citep[$e.g.$,][]{vincenzo19, belokurov20, feuillet21}.  
The GES merger affected the disk by heating it \citep[$e.g.$,][]{dimatteo19, belokurov20}, andhi may have also triggered the formation of the bar/bulge \citep{merrow24}.  

Tracing the contribution of GES in the inner Galaxy and at small Galactocentric distances is difficult.  Much of the halo close to the Galaxy's center is obscured by the prominent bar/bulge that dominates the local stellar population.  Although the bar/bulge tends to be more metal-rich than the halo, even when using metal-poor stars as tracers of the halo, the majority of metal-poor stars at $R <$ 2-3~kpc are confined to the bar/bulge \citep{kunder20, lucey21, ardernarentsen24}.  Therefore, even though it is thought that the GES contribution to the Galactic halo increases with decreasing Galactocentric distance \citep{iorio21}, this has not been able to be observationally confirmed within the distances in which the bulge/bar dominates.  Here we seek to search for the signature of GES in the stellar halo that surrounds the bar/bulge area using RR Lyrae stars (RRLs).

RRLs are widely adopted distance indicators with which to study the halo \citep[$e.g.$,][]{saha85, suntzeff91, liu22}.  The fundamental mode RRLs, RRab stars, pulsate radially with periods of $\sim$0.55~days and amplitudes of $\sim$1~mag in the optical; especially using RRab stars, clean samples of RRLs can and have been compiled.  However, the first-overtone RRc stars (with periods $\sim$ 0.35~days and amplitudes $\sim$0.3~mag in the optical) are also frequently used to enhance RRL datasets \citep[e.g.,][]{catelan15}.  RRLs are metal-poor and ubiquitous in the halo, and with an absolute magnitude of $M_V \sim$0.7~mag in the optical, they have been observed out to 200-300~kpc \citep[e.g.,][]{medina24, feng24}.  Here we use the sample of 8457 RRab and RRc inner Galaxy RRLs in \citet{prudil25a} to separate the bulge RRLs from the halo RRLs with the goal of understanding the contribution of GES in the stellar halo at small Galactocentric distances ($R \rm \lesssim 10~kpc$). 

This paper is part of ``The Galactic Bulge exploration" series, which has the goal of tracing the structure and dynamics of the inner Galaxy from the viewpoint of the old, metal-poor RRLs.  Our first papers set up modern techniques to determine RRL distances \citep{prudil24a}, radial velocities \citep{prudil24b}, and metallicities \citep{kunder24}, and our subsequent papers utilize these tools to probe the spatial distribution \citep{prudil25b} and dynamics \citep{prudil25a} of the Milky Way bulge.  This is our first paper in the series that focuses exclusively on the inner-central halo.

\section{The RR Lyrae Star Sample} \label{sec:sample}
\subsection{The halo RR Lyrae stars}
\begin{figure*}
\centering
\mbox{\subfigure{\includegraphics[height=5.3cm]{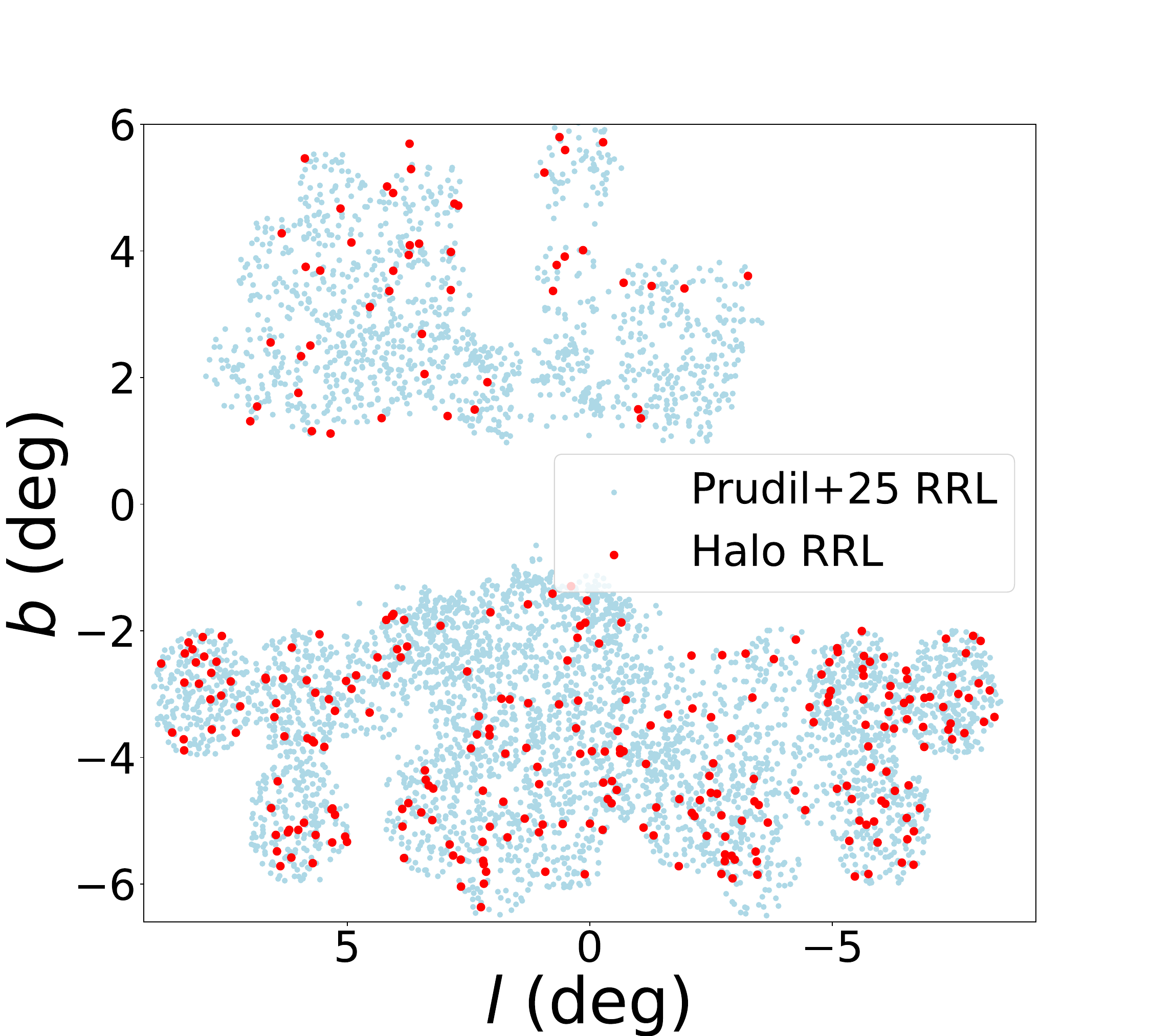}}}
\mbox{\subfigure{\includegraphics[height=5.3cm]{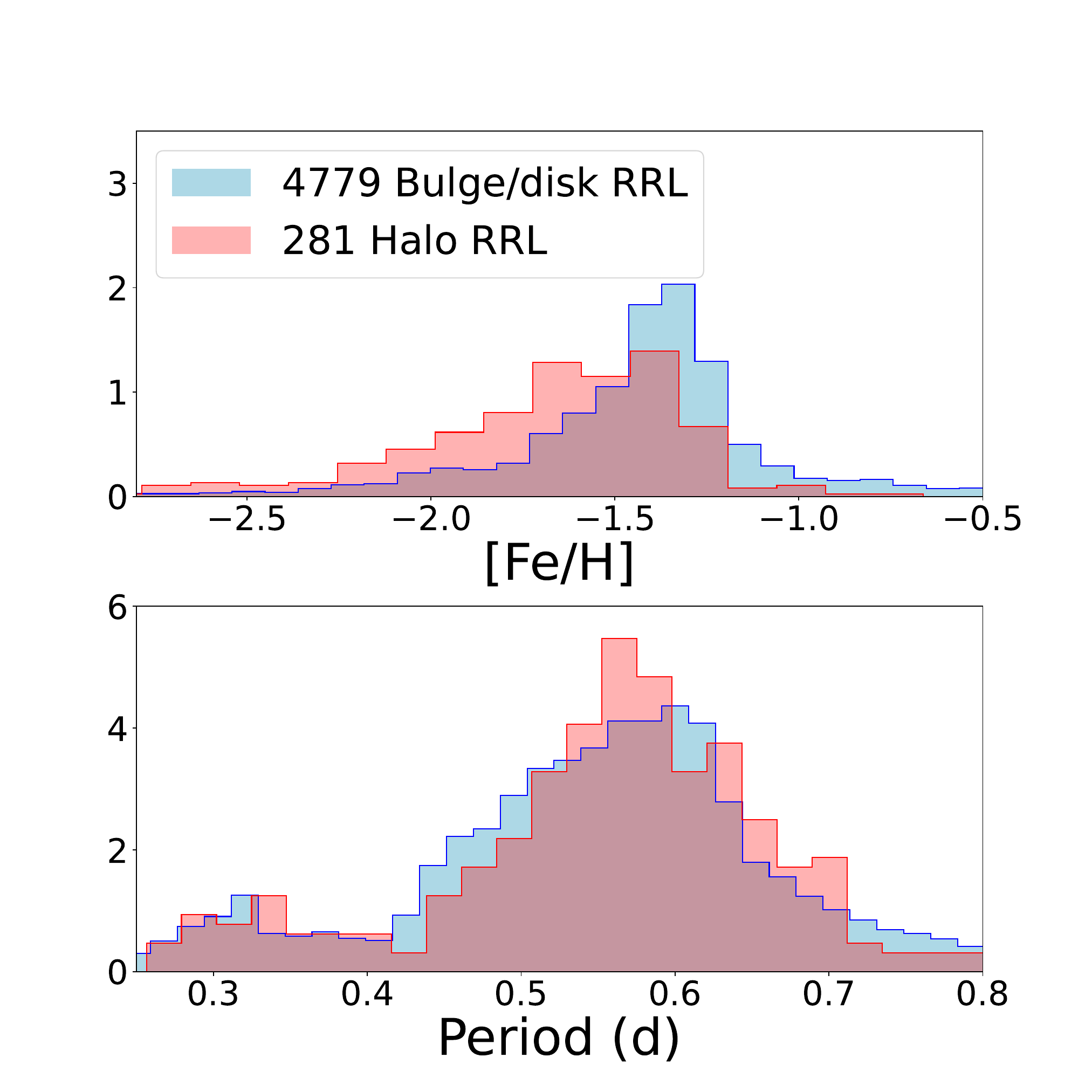}}}
\mbox{\subfigure{\includegraphics[height=5.3cm]{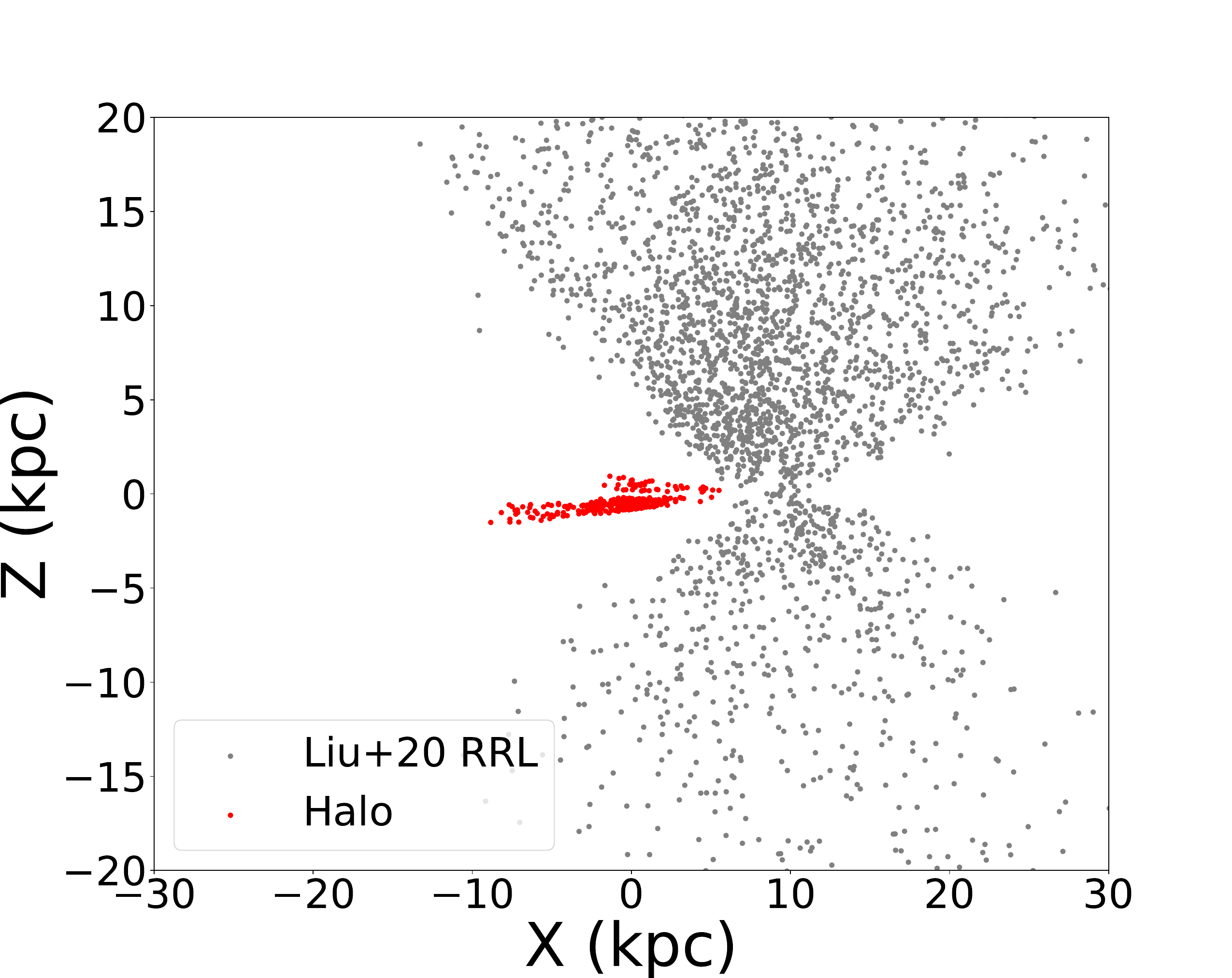}}}
\caption{
{\it Left:}. Our sample of RRLs in Galactic coordinates, where the RRLs with orbits consistent with belonging to the inner-central halo ($\rm 3~kpc < z_{max} < 20~kpc$) are highlighted. 
{\it Middle:}  The inner-central halo sample has a lower $\rm [Fe/H]$ metallicity and also a period distribution more consistent with accretion than the bulge/disk RRLs. 
{\it Right:} Our RRL sample $z-$coordinate plotted with $x-$coordinate in left-handed Cartesian Galactocentric coordinates. The grey and red dots represent the sample from \citet{liu20} and our sample, respectively.
}
\label{fig:sample}
\end{figure*}

We select RRLs from The Galactic Bulge Exploration 6D catalog \citep{prudil25a}, where the sample of stars is shown in Figure~\ref{fig:sample}.  Briefly, this is a collection of 8457 RRLs toward the inner Galaxy with all three components of velocity -- proper motions from {\it Gaia} DR3 and systemic radial velocity measurements derived using the \citet{prudil24b} radial velocity templates combined with spectra observed by APOGEE DR17, BRAVA-RR and/or the ESO archive PropID: 093.B-0473.  All these stars further have distances derived from newly calibrated period-luminosity-metallicity relations taking advantage of RRL parallaxes in {\it Gaia} DR3 \citep{prudil24a}.  To ensure the highest quality of data, only the 5108 stars with radial velocity uncertainties less than 10~km~s$^{-1}$ were used, as well as those having {\tt ruwe $<$1.4} and {\tt ipd\_frac\_multi\_peak < 5}. The {\tt ruwe} is the re-normalized unit weight error, and {\tt ipd\_frac\_multi\_peak} is a measure of the percentage of detection of a double peak in the point spread function (PSF) during {\it Gaia} image processing.  Lastly, we restrict the sample of RRLs to have $x$-distances of $-$10~kpc $< x <$ 10~kpc, to concentrate on the inner-central part of the Galaxy, which results in a sample of 5060 RRLs.  

The RRL positions and velocities are converted to the Galactocentric frame assuming a left-handed system with positive $U$ towards the Galactic anticenter, $V$ in the direction of Galactic rotation, and positive $W$ in the direction to the Galactic north pole. 
We adopt a Sun location of ($R,z,\phi$) = (8.1~kpc, 25~pc, 0) 
and assume a Sun's peculiar solar motion of ($U_{\odot}$, $V_{\odot}$, $W_{\odot}$) = ($-$11.1, 12.24, 7.25)~km~s$^{-1}$, 
and a local standard of rest velocity $v_{LSR}$ = 233~km~s$^{-1}$.  

In order to avoid stars belonging to the bulge, only stars on orbits consistent with belonging to an inner-central halo are selected.  
Stars in the Galactic bulge will be confined to the central part of the Galaxy, typically belonging to families of periodic orbits that support the bar/peanut shape of the ``bulge" \citep[e.g.,][]{portail15, prudil25a}.  In contrast, stars in an inner-central halo will not be confined to the central few kiloparsecs of the Milky Way, with orbits that are less organized, have higher velocity dispersions, and do not contribute to the 3D shape of the bulge/bar \citep[e.g.,][]{lucey22, cabreragarcia24}. 
Orbits are carried out using {\tt AGAMA}\footnote{http://agama.software/} \citep{vasiliev19} and adopting the potential from \citet{mcmillan17}.  This potential was selected to be consistent with other studies who have used 6D positions and kinematics to study GES \citep[e.g.,][]{helmi18, koppelman20, feuillet20}.
We define our inner-central halo star sample as those having $z_{max}$ between 3~kpc and 20~kpc, consistent with selecting halo stars from e.g., \citet{liu20} and \citet{cabreragarcia24}.  
There are 281 RRLs that fulfill these criteria.  The RRLs with $z_{max} <3$ belong primarily to the bulge, but those that also have large apocenter distances are consistent with being disk RRLs.

Figure~\ref{fig:sample} (middle panel) shows the periods and photometric $\rm [Fe/H]$ metallicities of our inner-central halo stars as compared to those from the bulge/disk.  These photometric metallicities are derived from the shape of the light curve of the RRL using relations from \citep{dekany21}, calibrated using the high-resolution \citet{crestani21} metallicities.  The typical uncertainty on these are 0.2~dex.

The RRLs with $z_{max} >3$ have lower metallicities and than those RRLs with $z_{max} <3$, in agreement with the inner halo being on average more metal-poor than the Galactic bulge \citep[e.g.,][]{zoccali17, lucey22}.  The double peak seen in periods of the fundamental mode halo RRLs (at $\sim$0.57 and $\sim$0.63 days) is in agreement with the halo harboring more accreted RRLs than the bulge/disk RRLs.  \citet{luongo24} show that the accreted RRL in the inner Galaxy have a bi-modal period distribution ($i.e.,$ exhibit the ``Oosterhoff" effect), whereas the $in~situ$ RRL exhibit a continuous period spread.  

The results of a two sided Kolmogorov–Smirnov (KS) test indicates that the RRLs with $z_{max} >3$ and those with $z_{max} <3$ trace different periods with KS-prob $< 0.01$ and also trace different photometric $\rm [Fe/H]$ metallicities with KS-prob $< 0.01$.  This KS-prob is less than the default threshold KS-prob = 0.05, below which one rejects the null hypothesis.  We therefore believe our cut at $z_{max} >3$ is effective in separating the bulge and disk RRLs from those belonging to the inner-central halo.  

Figure~\ref{fig:sample} shows our sample of halo RRLs compared to other surveys using RRLs to probe the halo from $e.g.$, the LAMOST Experiment for Galactic Understanding and Exploration (LEGUE) and the Sloan Extension for Galactic Understanding and Exploration (SEGUE) survey \citep{liu20}.  Our sample of 281 RRLs is unique in that it probes the stellar halo in a region where it is typically lost in the clutter of the Galactic bulge.  To our knowledge, these RRLs comprise the largest sample of Milky Way inner-central halo stars.

\subsection{Integrals of Motion}
The search for GES in the inner-central Milky Way can be complicated by the phase mixing of stars over time, 
especially since it arose in the early stages of the Milky Way formation.  
However, in a slowly evolving potential, N-body simulations indicate that dynamical coherency is retained in debris from the same progenitor in the space defined by the Integrals of Motion, $e.g.$, energy and angular momentum, $L_z$ \citep{johnston96, helmi00, gomez13}.  Observationally, \citet{helmi18} show that stars with slightly retrograde mean rotational motion of $-$1500 $< L_z <$ 150 $\rm kpc~km/s$ and energy $E > -$1.8 x 10$\rm ^5~km^2/s^2$ stand out as belonging to GES.  
This selection method will not produce a completely clean sample of GES stars, but has been shown to be useful in estimating some of the GES progenitor properties \citep[e.g.,][]{carrillo24}.  We use simulations of a GES-like merger in the following section, \S\ref{sec:auriga}, to estimate contamination in our sample and in this selection. 

\begin{figure}[h]
\centering
\mbox{\subfigure{\includegraphics[height=6.9cm]{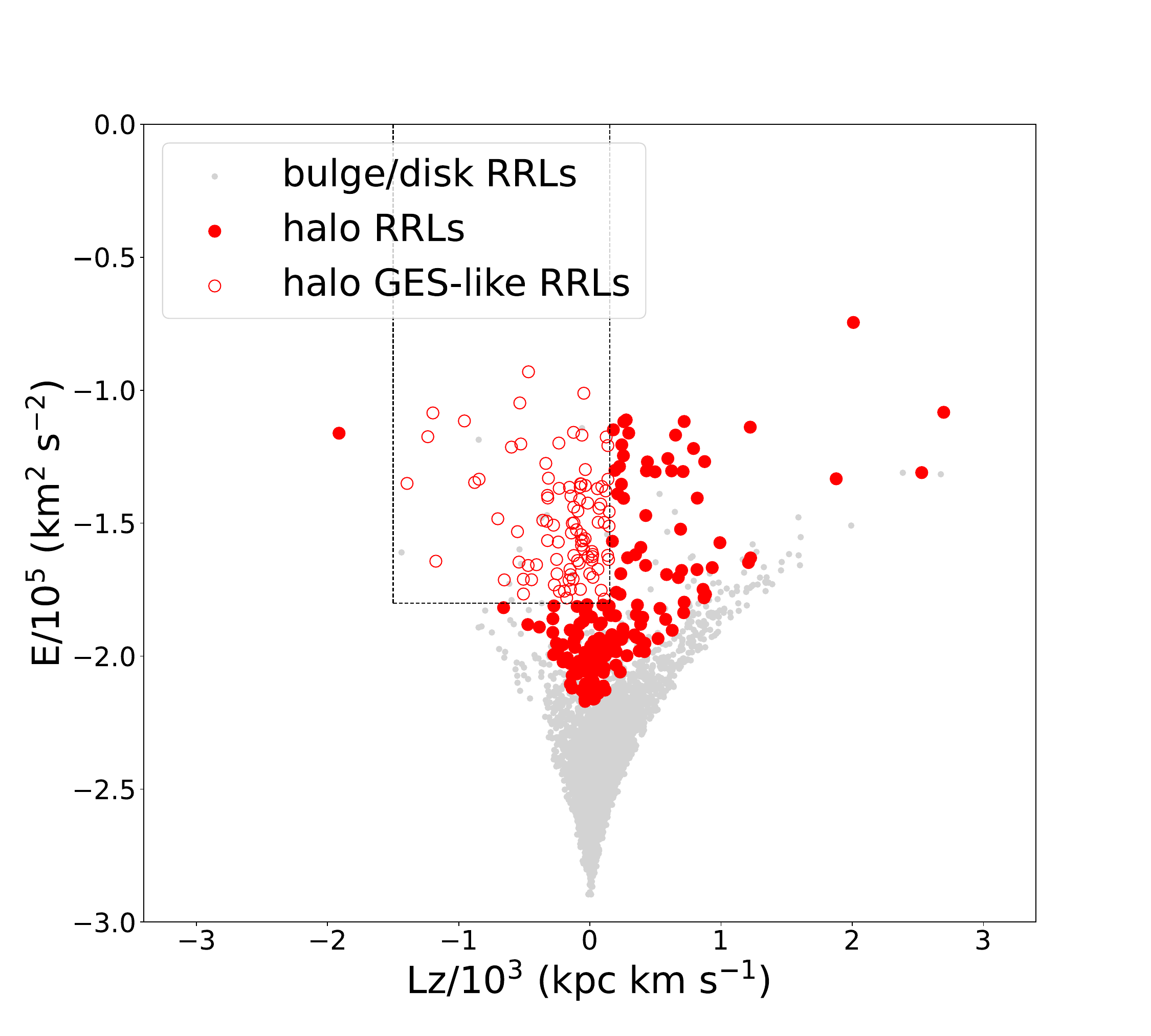}}}
\mbox{\subfigure{\includegraphics[height=6.9cm]{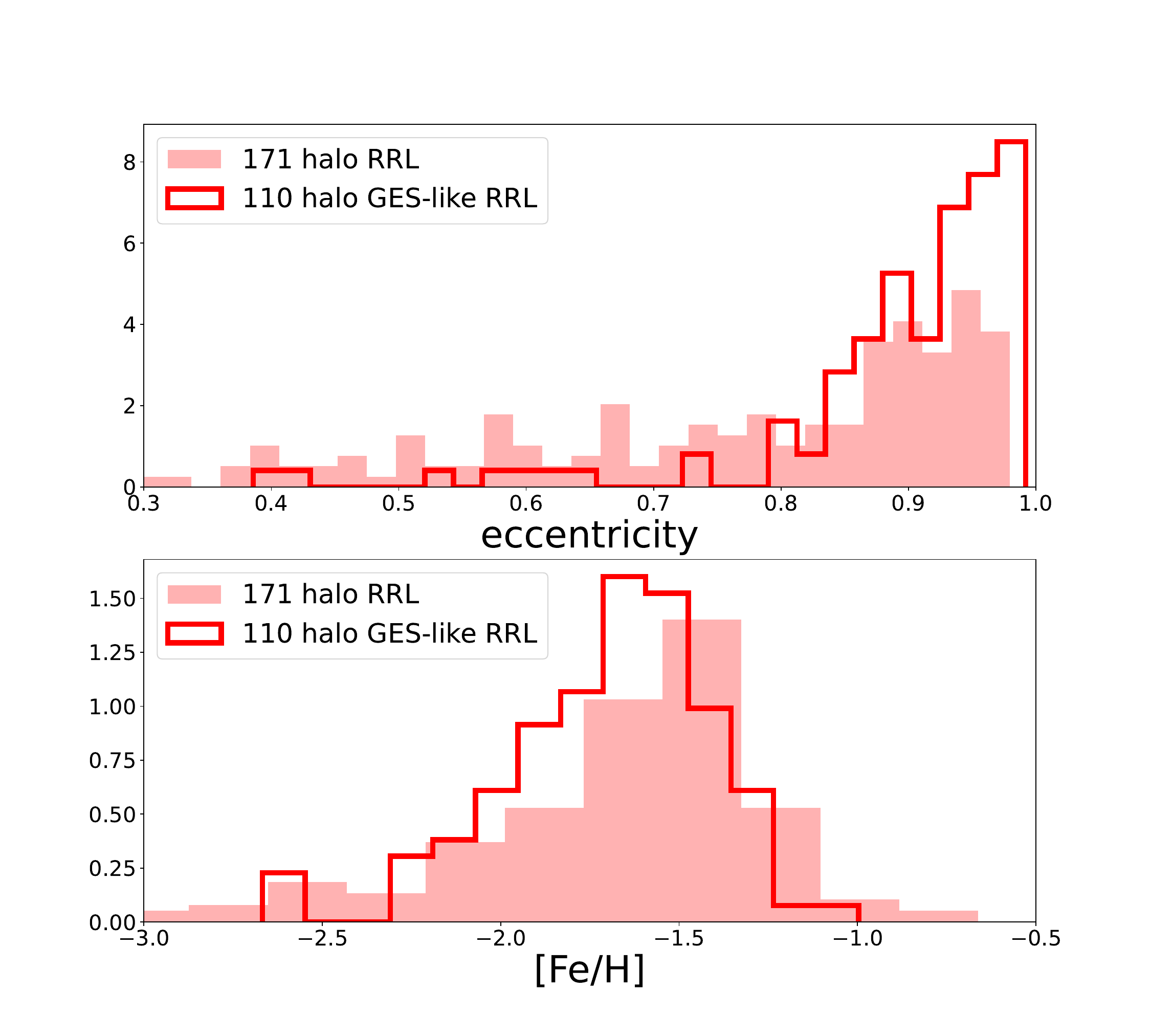}}}
\caption{{\it Top:} The inner Galaxy RRL sample in $L_z$ and energy space.  The inner-central halo RRLs are designated with large red circles, whereas the bulge and disk RRLs are shown as small grey circles.  The straight lines indicate the criteria used to select GES stars, namely $-$1500 $<  L_z <$ 150 kpc~km~s$^{-1}$ and $E > -$1.8 x10$^5$ km$^2$~s$^{-2}$.  
{\it Bottom:}. The RRL with $L_z$ and energy values consistent with GES have higher eccentricities and tend to be more metal-poor as compared to other inner-central halo RRLs.
}
\label{fig:RRL_Lz_E}
\end{figure}

We divide our sample into two groups -- one with the angular momentum and energy values that GES stars tend to exhibit, and the other group of stars with integrals of motion that fall outside those $L_z$ and energy values.  Figure~\ref{fig:RRL_Lz_E} shows that the resulting {\it Gaia}-Enceladus-Sausage and non-GES groups contain $110$ and $171$ RRLs, respectively.  
The RRLs with the angular momentum and energy values consistent with GES have high eccentricities and tend to also be more metal-poor than the other halo RRLs.  Stars with high eccentricities ($e >$0.75) are one of the telling indicators of a GES star \citep[e.g.,][]{carrillo24, myeong22} and the clear peak of halo GES-like RRL with high eccentricities indicates at least some contribution of GES to the inner-central halo.  

The GES RRL metallicity distribution function (MDF) shown here will be different than other investigations of the MDF of GES because our RRL sample is only dependent only on stars residing the instability strip of the horizontal branch (HB). HB morphology changes as a function of metallicity and the corresponding change in the population of the instability strip.  There is a failure of the metal-rich red HB stars to penetrate into the instability strip, and therefore a greater frequency of metal-poor stars become RRL variables.  This is typically why RRLs compromise a more metal-weak tail of a population \citep[e.g.,][]{lee92, catelan15}.  The MDF shown here is similar to other studies that probe the metal-weak tail of GES; in particular, both the RRLs and the metal-weak tail of GES seen in \citet{bonifacio21} 
show a broad $\rm [Fe/H]$ peak at  $\rm [Fe/H] \sim -$1.5.

\section{Auriga Simulation} \label{sec:auriga}
Although a number of studies have shown that accreted stars from GES have a preponderance for the energies and angular momenta in the regime probed here, both $in~situ$ and accreted stars can posses such energies and angular momenta \citep[e.g.,][]{ceccarelli24, feltzing23}.  
Here we turn to simulations to help estimate the likelihood that the RRLs with specific energies and angular momentum originated from GES.
In particular, the Auriga simulations \citep{grand17} are a set of 
30 cosmological magneto-hydrodynamical zoom simulations of the formation of galaxies in isolated Milky Way mass dark halos.
These simulations, that are publicly available \citep{grand24}, have been shown to reproduce a wide range of phenomena observed in Milky Way-mass galaxies, such as the sizes, rotation curves, star formation rates, and metallicities.  Of the different Auriga
simulations, the Auriga 18 (Au-18) simulation has a quiescent merger history in line with that of the Milky Way, with chemodynamics especially similar to that of the Milky Way \citep[e.g.,][]{fragkoudi20}. The Au-18 simulation further has a merging dwarf galaxy with properties ($e.g.,$ mass, time of merger) comparable to that of the GES in the Milky Way, dubbed GES-18 \citep{merrow24}.  

Stellar particles in Au-18 with the same Galactic latitude and longitude coordinates as our RRL sample are selected.  We further require the particles to have $y$-distances of $-$4 $< y <$ 4~kpc and $z$-distances between $-$3.5 $< z <$  3.5~kpc, ($i.e.,$ Figure~\ref{fig:sample}), as well as ages older than 10~Gyr and metallicities more metal-poor than $\rm [Fe/H] <-$0.5.  This produces a catalog of 94,066 particles -- 81,687 having formed $in~situ$ and 12,379 being accreted.  
Orbits are calculated for those particles using a potential constructed in the {\tt galpy}{\footnote{\tt http://github.com/jobovy/galpy}} Python package \citep{bovy15} that reconstructs the total fitted mass profile of Au-18 at $z=$0.  The adopted halo virial mass, halo virial radius, stellar mass, stellar disk mass, radial scalelength, inferred stellar bulge mass, and bulge effective radius for the Au-18 potential comes from Table~1 in \citet{grand17}.
We adopt as our Au-18 inner-central halo sample the 8592 particles with Galactic coordinates consistent with the RRL sample ($i.e.$, $ |l| <  9^\circ$ and $1^\circ < |b| <~$6.5$^\circ$, see Figure~\ref{fig:sample}), as well as particles with 20~kpc~$> z_{max} >$3~kpc and $x$-distances of $-$10~kpc $< x <$ 10~kpc.  

\subsection{GES contribution using eccentricity as an indicator}

\begin{figure*}
\centering
\mbox{\subfigure{\includegraphics[height=5.2cm]{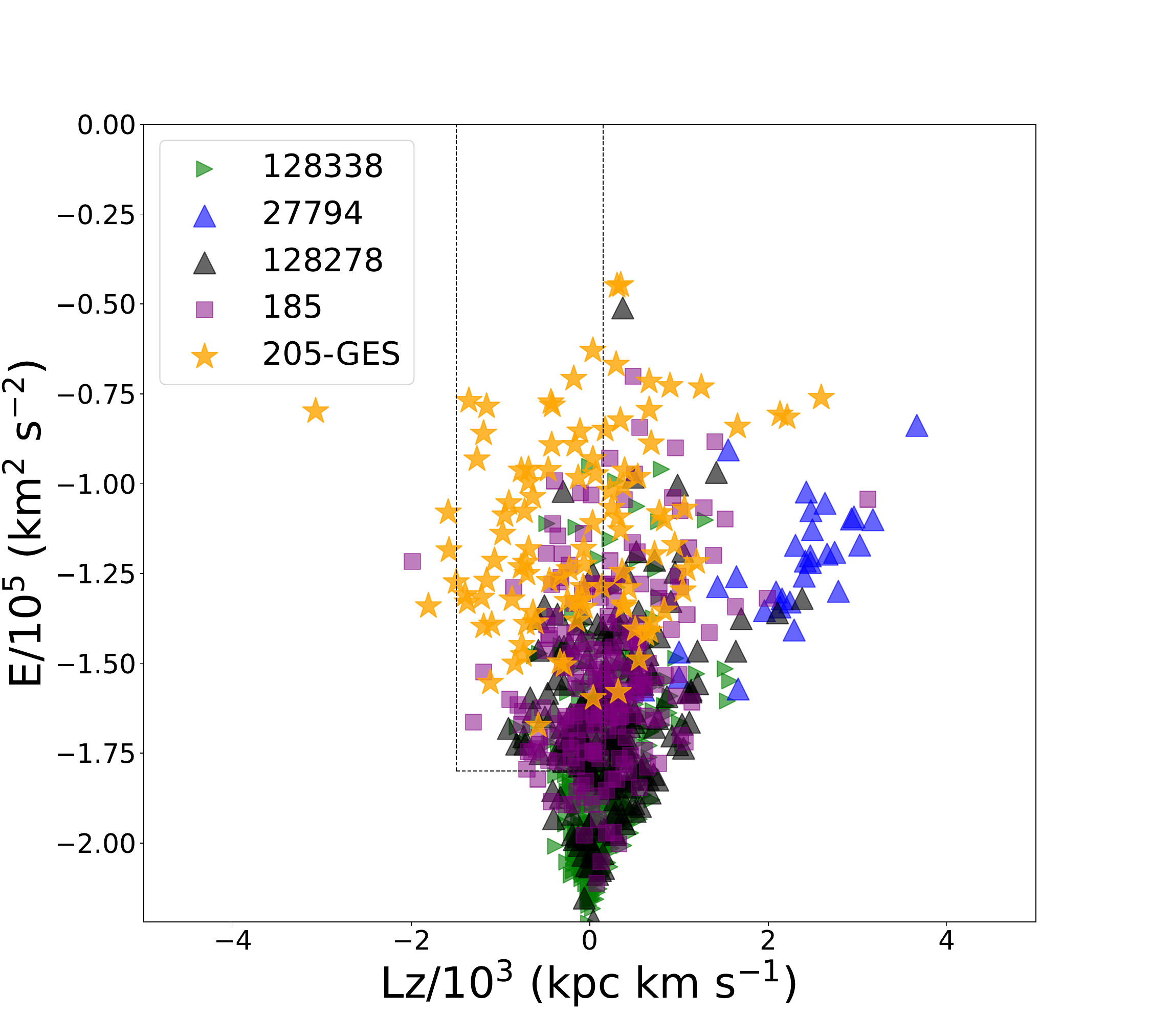}}}
\mbox{\subfigure{\includegraphics[height=5.2cm]{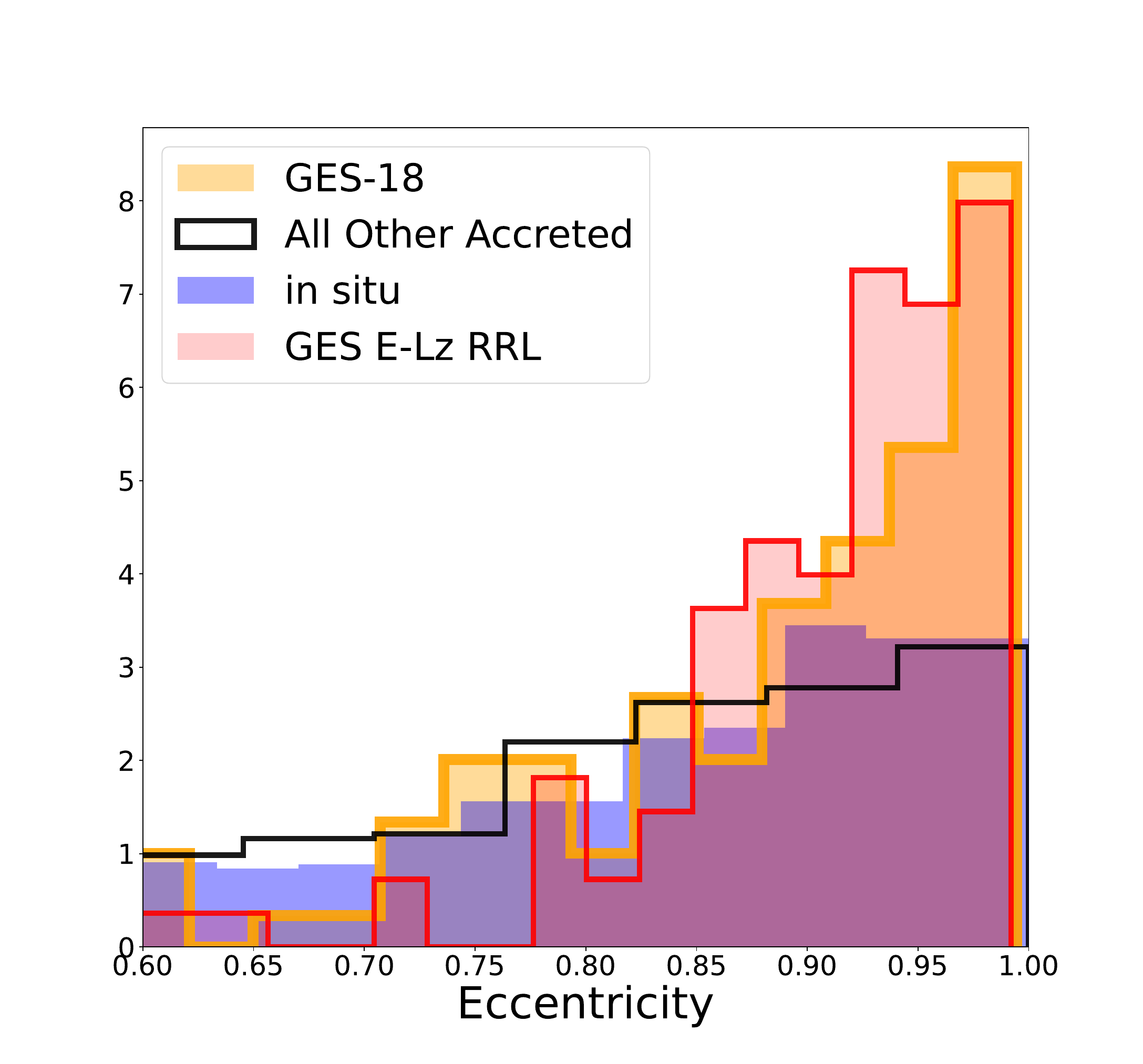}}}
\mbox{\subfigure{\includegraphics[height=5.2cm]{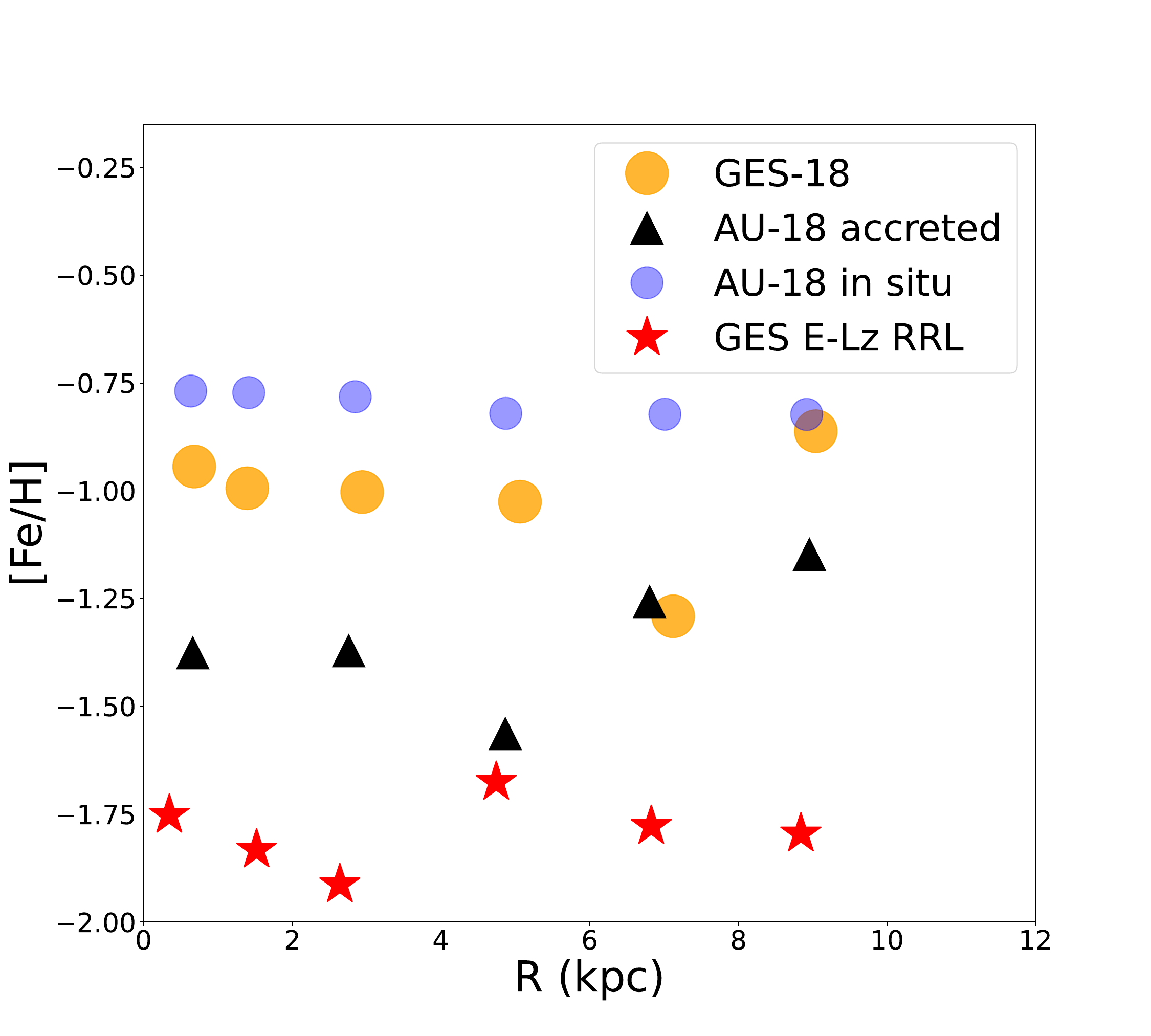}}}
\caption{
{\it Left:} The integrals of motion for five dominant accretion events in the Au-18 inner-central halo.  The \citet{helmi18} Lz-E criteria to select GES stars encompasses a large fraction of the GES-18 ({\tt peak\_mass\_ID=}205) particles (rectangle region).
{\it Middle:} The distribution of eccentricity of GES-18 particles as compared to the $in~situ$ particles and the other accreted particles in the inner-central halo of Au-18.  The eccentricities of the observed halo RRLs with integrals of motion consistent with GES are also shown.  GES-18 dominates at high eccentricities; the peak of high eccentricities in our sample of halo RRLs suggest $\sim$9\% of the sample originated from the GES merger.
{\it Right:} Galactocentric radius versus $\rm [Fe/H]$ metallicity for GES, the other accreted components, the $in~situ$ components and the RRLs in the inner-central halo.  Only the $in~situ$ Au-18 particles exhibit a statistically significant negative metallicity gradient with Galactocentric radius, with a slope of $-$0.0075 $\pm$ 0.0015 dex~kpc$^{-1}$. 
}
\label{fig:ELz_auriga}
\end{figure*}

Figure~\ref{fig:ELz_auriga} (left panel) shows the position of five of the more massive accreted satellites present in our inner-central Au-18 sample in energy and angular momentum space.  The simulated GES analog, GES-18, although a dominant accretion event in the solar vicinity, is not necessarily the most massive accreted remnant in the inner-central halo.  In this Auriga simulated MW-analog, a few satellites that accreted earlier in the history of the formation of the Galaxy ($e.g.$, with {\tt peak\_mass\_ID=}128338, 128278 and 185) contributed more mass to this part of the Galactic halo.  The {\tt peak\_mass\_ID} is the Auriga identification given to the simulated satellite galaxy when it reaches its maximum mass in an Auriga simulation.  
The structures that accreted earlier in the history of the Milky Way have lower energies, and some accreted satellites have integrals of motion that are similar to the disk (e.g., those with {\tt peak\_mass\_ID=}27794).  GES-18 (with {\tt peak\_mass\_ID=}205) indeed consists of particles with energy and angular momentum that primarily encompass the regime first used by \citet{helmi18} to detect the signature of GES.  The correspondence between the observed integrals of motion associated with GES and those of the simulated GES analog suggests good agreement between the models and observations in reproducing the observed dynamics.  
We find that only a handful of the GES-18 particles have $z_{max} < 3~\rm kpc$ (see $e.g.,$ Figure~\ref{fig:RRL_Lz_E}), indicating that there should be little to no debris from GES in bulge RRL samples selected by culling the bulge from the halo using orbital parameters.

The middle panel of Figure~\ref{fig:ELz_auriga} shows the eccentricities of the GES-18 particles as compared to the eccentricities of the $in~situ$ particles, and those from the other accreted events, where the eccentricity is determined from the orbital integration described above.  In particular, eccentricity is defined as $e = (r_{apo} - r_{peri})/(r_{apo} + r_{peri})$ where $r_{apo}$ and $r_{peri}$ are the apocenter and pericenter distance of the Galactic orbit.  The eccentricities of the observed 281 RRL sample is also shown.  GES-18 particles dominate in the high eccentricity end, with $\sim$70\% of the inner-central halo GES-18 particles having eccentricities $>$ 0.85.  In contrast, only 42\% of the $in~situ$ particles and 44\% of the accreted particles have eccentricities $>$ 0.85.  The RRL sample has a clear peak of stars at high eccentricities -- 51\% of the inner-central RRLs have eccentricities $>$ 0.85.  This is an $\sim$9\% increase to what is expected from either an $in~situ$ population or an accreted population devoid of a GES-like merger.  Therefore, from a comparison of the eccentricities of the RRL sample to the Au-18 simulation, the excess of high eccentricity RRLs suggests $\sim$9 $\pm$ 2\% of these inner-central halo RRLs originated from GES.  The uncertainty on the fraction is derived from Poisson statistics, assuming the error arises from the sample size of the RRLs with high eccentricities.

\subsection{GES Radial Metallicity Gradient}
Dwarf spheroidal galaxies (dSphs) orbiting the Milky Way typically show a negative metallicity gradient with radius \citep[see, e.g.,][]{tolstoy23}.  The center of these galaxies have higher gas densities, and therefore undergo a more intense star-formation-rate as compared to their outer regions of the dSph galaxy.  
Chemical gradients in merger debris have been found for Sagittarius, as its core is $\sim$0.7~dex more metal-rich as compared to its streams \citep[e.g.,][]{alard01, bellazzini99, hayes20, minelli23}. 

By using RRLs in the inner halo, \citet{liu22} report a tentative signature of a negative metallicity gradient in the LAMOST RRLs of GES when looking at radial distances between $\rm 7 < R < 27$~kpc.  
\citet{medina25} find that if there is a metallicity gradient, the metallicity gradient is a factor of two smaller than reported by \citet{liu22}.  
Within our inner-central RRL halo sample, we find no clear metallicity gradient between $\rm 0 < R < 9$~kpc.     
Figure~\ref{fig:ELz_auriga} (right panel) shows the $\rm [Fe/H]$ metallicity of the RRLs with integrals of motion consistent with GES as a function of Galactocentric distance.  
The Auriga simulation of GES-18 also does not indicate a radial metallicity gradient with respect to the Galactic center in the inner-central halo.  
The only possibly significant radial metallicity gradient is in the $in~situ$ Au-18 particles, which exhibit a slope of $\rm -0.007 \pm 0.002~dex~kpc^{-1}$.  
The negative metallicity gradient expected for $in~situ$ stars in the inner Galaxy, based on tracing stars formed $in~situ$ in the Au-18 simulation, indicates that observed negative metallicity gradients in the Milky Way could arise from contamination with $in~situ$ stars in a GES sample.
The absence of a radial metallicity gradient in GES indicates that either the central core of GES is not dominant in the inner-central halo of the Milky Way and/or that the chemical evolution in this GES merger is relatively well-mixed \citep{ciuca23}.

\subsection{GES contribution using integrals of motion as an indicator}
\begin{figure*}
\centering
\mbox{\subfigure{\includegraphics[height=5.2cm]{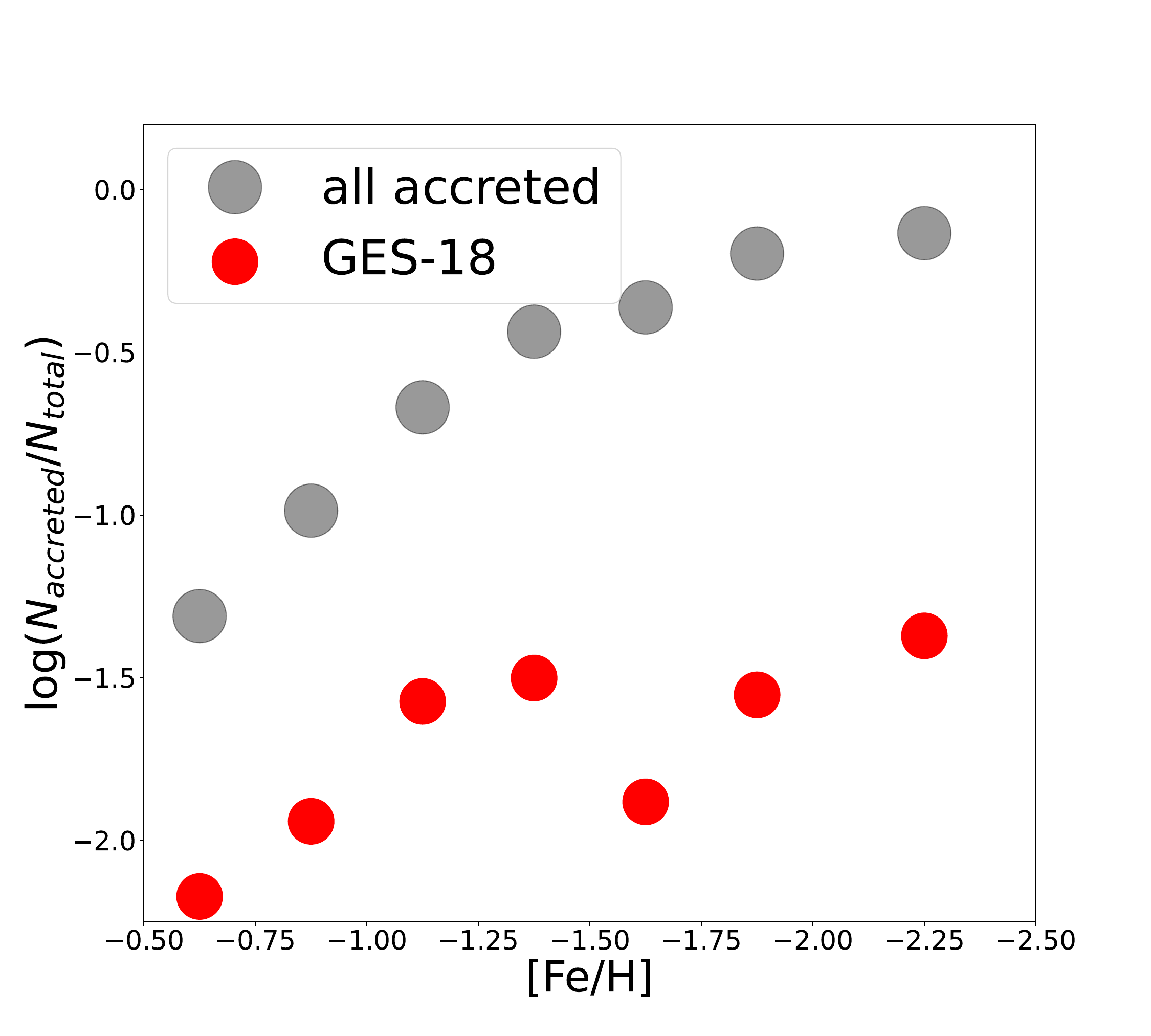}}}
\mbox{\subfigure{\includegraphics[height=5.2cm]{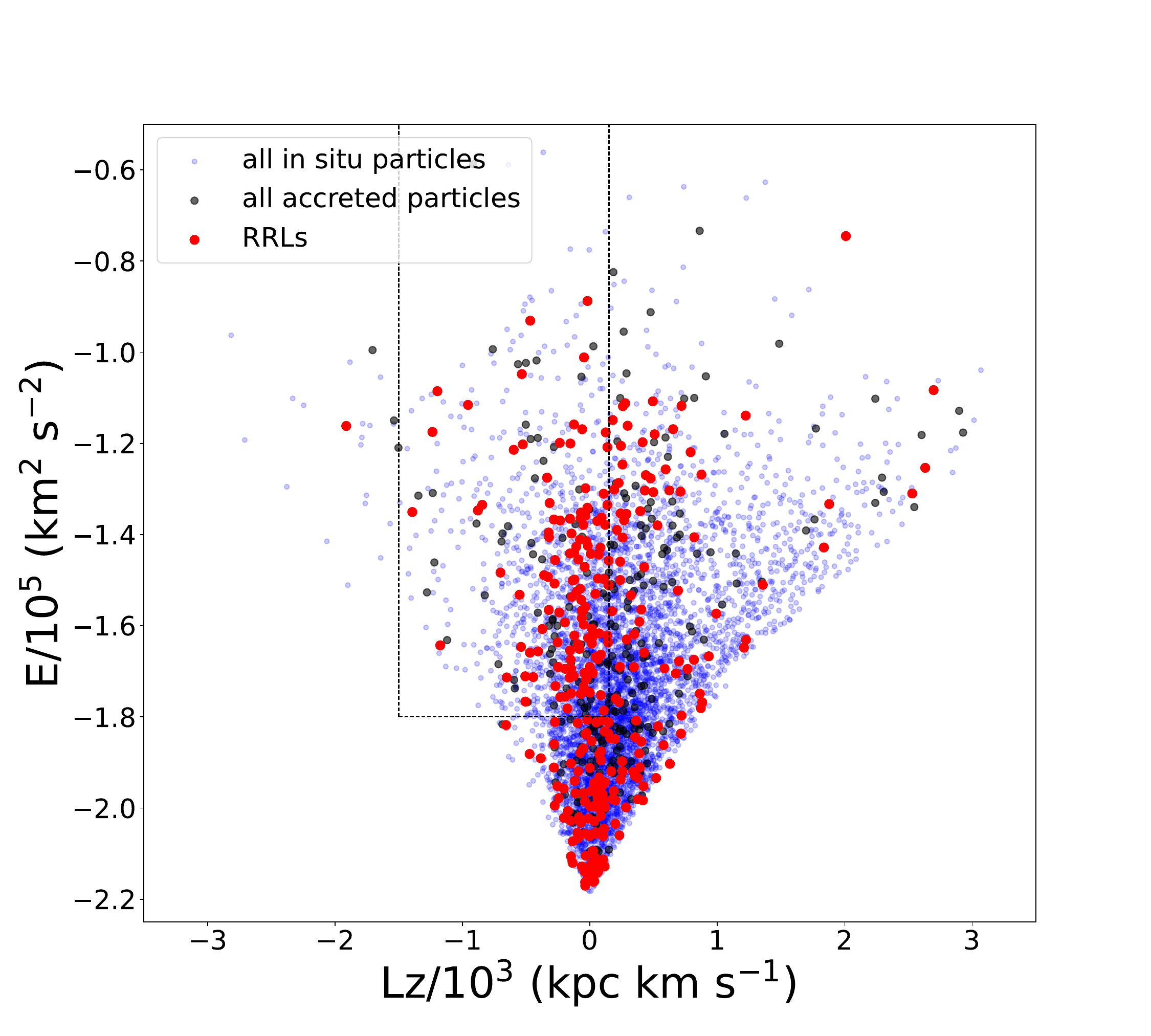}}}
\mbox{\subfigure{\includegraphics[height=5.2cm]{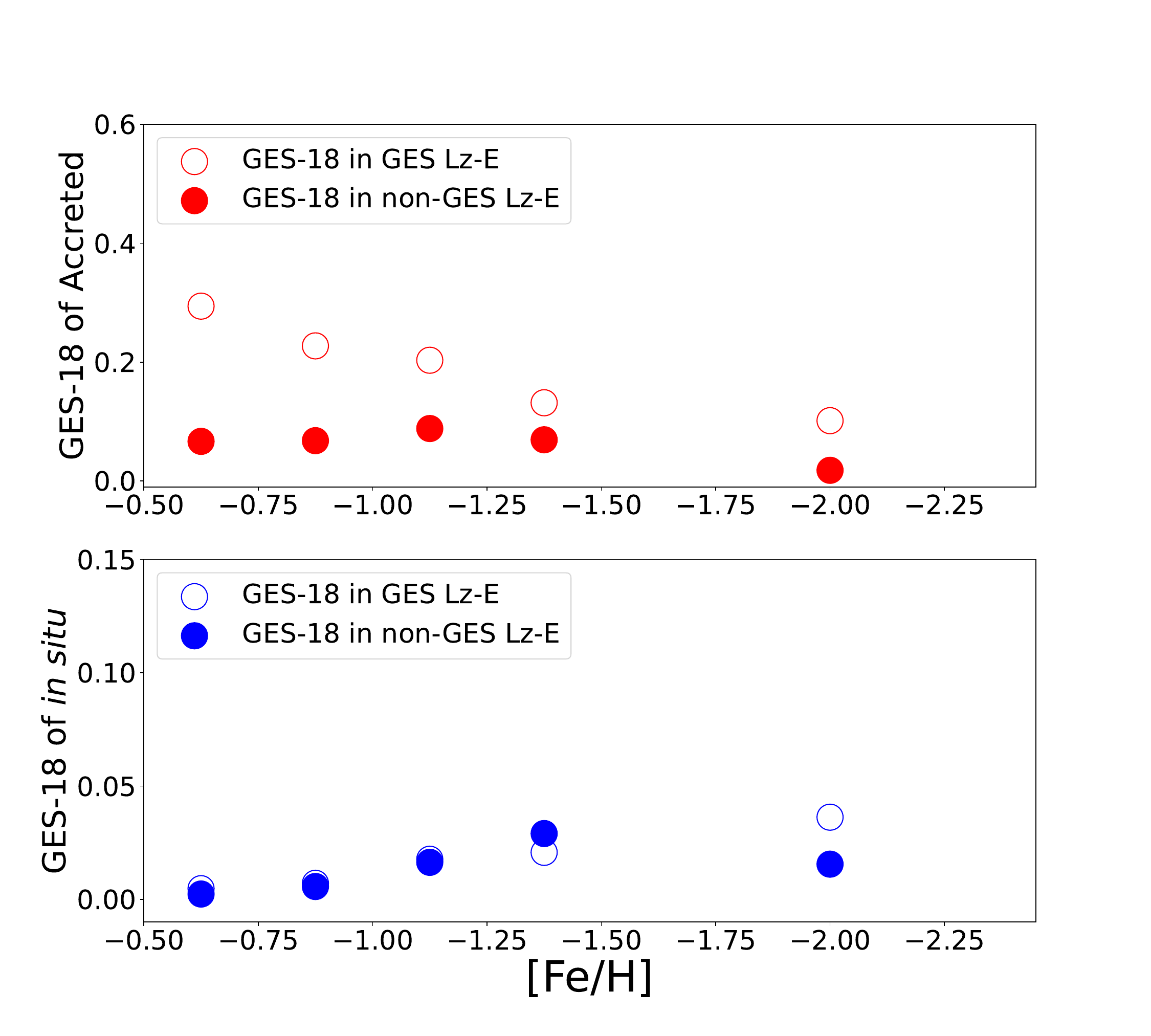}}}
\caption{
{\it Left:}  The logarithm of the fraction of the total accreted particles (grey) and GES-18 particles (red) in the inner-central halo from the Au-18 simulation. The fraction of GES-18 merger particles peaks at $\sim -$1.3~dex, whereas there is a dramatic increase with decreasing metallicity for other accretion events. 
{\it Middle:}  The integrals of motion for the Au-18 inner-central halo particles compared to the RRL sample.  The \citet{helmi18} Lz-E criteria to select GES stars is indicated by the dashed line.
{\it Right:} 
The fraction of GES-18 particles as compared to the Au-18 accreted particles (top) and the fraction of GES-18 particles as compared to the Au-18 $in~situ$ particles (bottom) that fall within the \citet{helmi18} Lz-E criteria is shown as a function of $\rm [Fe/H]$ metallicity.  The fraction of GES-18 particles that falls outside the \citet{helmi18} Lz-E criteria is shown as a function of $\rm [Fe/H]$ metallicity.  In samples of accreted stars, $\sim$30\% of stars in the \citet{helmi18} Lz-E criteria will belong to GES.
}
\label{fig:contamination_auriga}
\end{figure*}

The MW-analogue, Au-18, can provide information on the fraction of accreted stars in simulated galaxies.  
In our selected inner-central region of Au-18, 13\% of the particles are accreted, but the fraction of accreted particles to $in~situ$ particles is strongly dependent on $\rm [Fe/H]$ metallicity.  
This is illustrated in the left panel of Figure~\ref{fig:contamination_auriga}, which shows the fraction of accreted particles compared to the $in~situ$ particles.  In the inner-central halo, the fraction of accreted stars increases significantly in more metal-poor populations.  At around $\rm [Fe/H] \sim -$1.5 there are roughly as many stars in the inner-central halo that were accreted as were formed $in~situ$, and at metallicity of $\rm [Fe/H] \sim -$2.0, 80 per cent of stars originated from accretion.  The fraction of GES-18 particles in the inner-central halo peaks at a metallicity of $\rm [Fe/H] \sim -$1.3.  GES was one of the more massive galaxies to merge with the Milky Way, and so its population of stars were more efficient in chemical enrichment and have higher fractions of stars in the more metal-rich end as compared to the bulk of the accreted population.  That GES stars have a peak in metallicity between $\rm [Fe/H] \sim$ $-$1.0 and $-$1.4 is consistent with metallicity distribution functions in $e.g.,$ \citet{myeong19, feuillet20}.

The middle panel shows the RRL integrals of motion compared to the $in~situ$ and accreted Auriga particles in the inner-central halo.  This implies many $in~situ$ particles will have integrals of motion that overlap with the angular momenta and energies of GES stars.  In agreement with what is seen in the MW \citep[e.g.,][]{das20}, the fraction of $in~situ$ to accreted particles decreases significantly when moving to lower $\rm [Fe/H]$ metallicities.

We seek to estimate the fraction of the $in~situ$ and other accreted Auriga particles in the integral of motion space in which GES dominates. 
The right panel of Figure~\ref{fig:contamination_auriga} shows the fraction of GES-18 particles compared to both $in~situ$ and other accreted particles.  Here, the particles are separated based on those with integrals of motion that fall in the \citet{helmi18} GES regime, and those that do not.  Especially at metallicities more metal-rich than $\rm [Fe/H] \sim-$1.6, the \citet{helmi18} selected integrals of motion do help distinguish GES stars from other accreted stars.

Turning to our RRL observations, 39\% of the inner-central halo RRLs have $L_z$ and energies values consistent with GES, and $\sim$25\% of these will be from GES as compared to other accreted events.   If 40\% of the inner-central halo RRLs were accreted (from left panel Figure~\ref{fig:contamination_auriga}), then roughly $\sim$4\% of RRLs in the \citet{helmi18} GES regime originated from GES.
Of the 61\% inner-central halo RRLs that fall outside the $L_z$ and energy values consistent with GES, $\sim$10\% will be from GES as compared to other accreted events. Again assuming 40\% of the inner-central halo RRLs were accreted, this means that roughly $\sim$2\% of RRLs in the non-GES regime originated from GES.  
Taking into account potential contamination from both the $in~situ$ population and other accreted debris in the inner-central halo—as suggested by the Auriga simulations—the estimated fraction of inner-central halo RRLs that may have originated from GES is approximately 6 $\pm$ 2\%.  The uncertainty on the fraction is derived from Poisson statistics. 
This is a slighter lower estimated fraction of GES RRLs than the $\sim$9\% estimation using eccentricity arguments above. 

%


\section{Discussion}
The large fraction of inner-central halo RRLs with angular momentum and energy values that overlap with GES ($\sim$39\% of the RRLs), and the peak of RRLs at high eccentricities, indicates that at least part of the RRLs in the inner Galaxy originated from GES.  
In comparison with the Auriga simulation that includes a GES analogue (Au-18), it appears 
the integrals of motion \citet{helmi18} used to first discover GES can be useful to disentangle GES from the field, but there will still be contamination from both $in~situ$ and other accreted stars.  Comparing the energy-$L_z$ values of the RRL sample with the integrals of motion that arise from the $in~situ$ and other accreted particles in Auriga, $\sim$6\% of the inner-central RRL originated from GES.  This fraction is sensitive to the $\rm [Fe/H]$ metallicity distribution of the RRLs, as the contamination of $in~situ$ stars drastically decreases with decreasing metallicity.

The Au-18 simulation shows that high eccentricities naturally arise from particles belonging to GES-18 in the inner-central halo, but that $in~situ$ and other accreted particles do not tend to have such high fractions of high eccentricity values.  The RRL sample has $\sim$9 $\pm$ 2\% more stars at eccentricities $>$ 0.85 than can be attributed to either $in~situ$ or other accreted stars.  This suggests that the peak of inner-central halo RRL with high eccentricities originated form GES, and therefore that $\sim$9 $\pm$ 2\% of the inner-central RRL originated from GES.  

\begin{figure}[hb]
\centering
\mbox{\subfigure{\includegraphics[height=6.2cm]{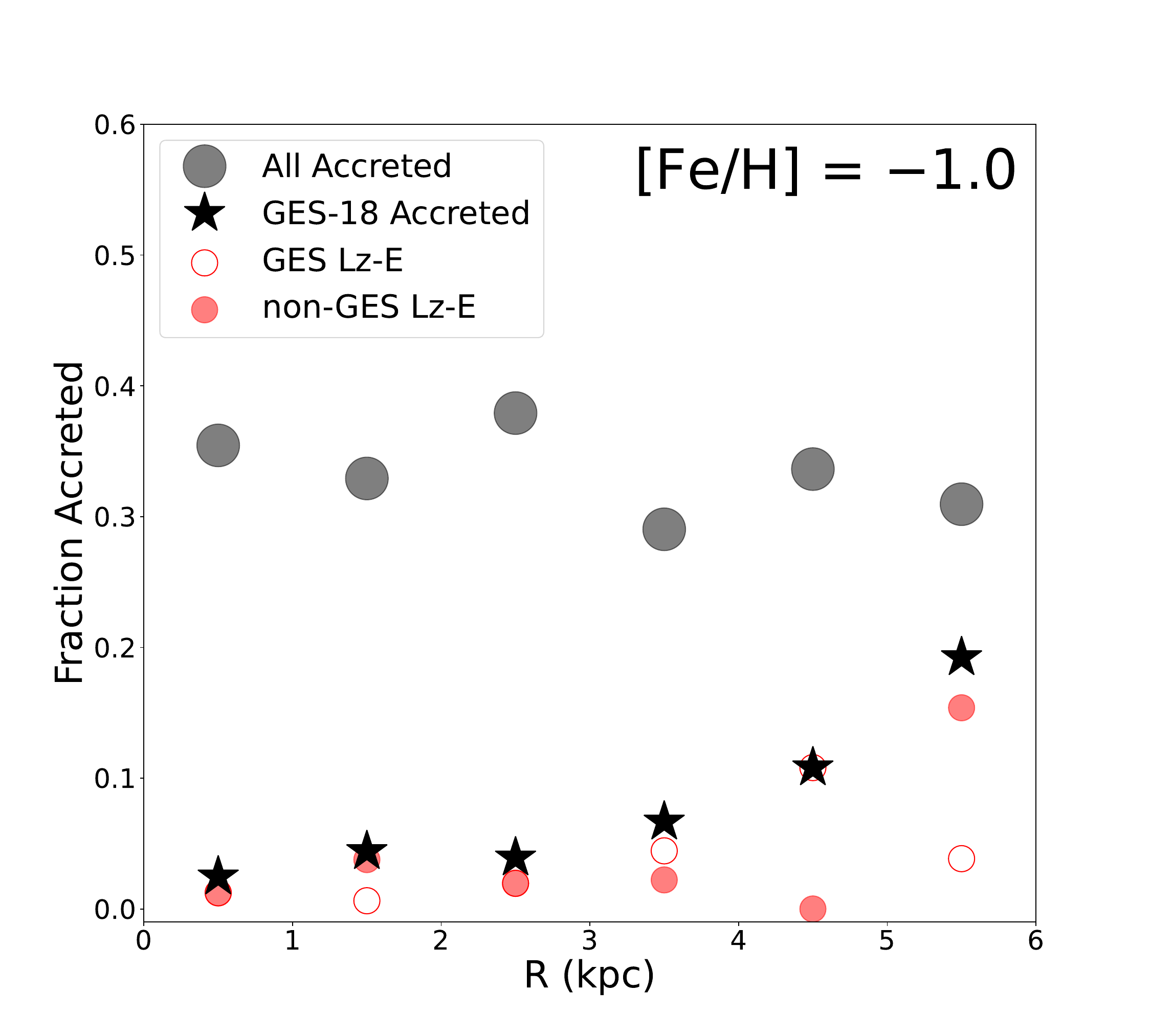}}}
\caption{The fraction of accreted particles with $\rm [Fe/H] = -$1 as a function of Galactocentric radius from the Au-18 simulation.  Although the fraction of accreted particles increases with increasing distance to the Galactic center, the opposite trend is seen for the GES-18 particles.   
}
\label{fig:Rfrac}
\end{figure}

Previous results suggest that $\sim$25-30\% of the RRL in the solar vicinity originated from GES \citep{prudil20, zinn20}, which is a considerably larger fraction than the RRL signature of GES in the inner-central halo.  
We can turn to the Auriga simulation to examine the density distribution of GES debris within the simulated stellar halo.  
Figure~\ref{fig:Rfrac} shows the fraction of accreted stars with $\rm [Fe/H] < -$1.0 as a function of Galactocentric radii.  
Whereas there is a higher fraction of accreted debris at small Galactocentric distances in the inner-central halo, this is not the case for GES-18.  GES-18 is not as dominant at small Galactocentric distances, instead becoming a larger fraction of the accreted population at Galactocentric distances of $R_{GC}\sim$5-6~kpc.  

GES analogue's therefore predict the contribution of GES being larger in the solar vicinity as compared to the densely populated inner-central halo, in-line with the RRL results presented here.  
Earlier accretion events had more time to fall into the potential well of the inner-central region, and more massive accretion events will sink more rapidly to the center due to dynamical friction \citep{amorisco17}.  

Such a trend has also been seen by \citet{naidu19} using the H3 Spectroscopic Survey.  In particular, they find that the relative fraction of GES stars decreases from $\sim 70\%$ at $R_{gal} \sim$20~kpc to $\sim 25\%$ at $R_{gal} \sim$6~kpc.  Their relative GES fraction compared to other Milky Way structures of $\sim$25\% at $R_{gal} \sim$6~kpc is similar to the  GES fraction of $\sim 20\%$ at $R_{gal} \sim$5~kpc found using Au-18.

\section{Conclusions}
The stellar halo in the inner-central regions of the Milky Way is largely unexplored, as the vast majority of studies of the Galaxy's stellar halo target stars along sight lines at high galactic latitudes, where there is significantly less contamination from stars residing in the disk and/or bulge \citep[e.g.,][]{conroy19, liu20}.  
However, simulations suggest that stars from early accretion events are to be preferentially found in the inner regions of haloes, while the outer parts are in general dominated by late accretion events \citep[e.g.,][]{zolotov09, tissera13}.
Probing the stellar halo in the inner-central regions of the Milky Way is therefore important in gaining the full picture of the accretion history of the Galaxy.

RRLs have long been used as probes of the stellar halo.  Analyzing a sample of 8457 RRL stars toward the Galactic bulge from \citet{prudil25b}, we identify 281 RRLs with kinematics indicating they are not confined to the bulge, but are instead part of the inner-central halo.  We compare the observed chemo-dynamical properties of the RRLs to a cosmological zoom-in simulation of a Milky Way-like analogue from the Auriga suite, Au-18, which suggests that 
$\sim$50\% of the RRL population in the inner-central halo is consistent with originating from accretion (Figure~\ref{fig:contamination_auriga}).  

Isolating particles from the GES analogue in Au-18 simulation that reside in the inner-central halo, 
we find that GES-18 particles cluster in a similar Energy-$\rm L_z$ plane as first reported by \citet{helmi18}.  
However, there are other merger remnants, as well as $in~situ$ stars, that have such integrals of motions, so energy and angular momentum alone does not allow a clean sample of GES to be selected (Figure~\ref{fig:ELz_auriga}, left panel).  Comparing the observed RRL integrals of motions with the particles in Au-18 and GES-18, we estimate $\sim$6\% of the RRLs in the inner-central halo originated from GES.
GES-18 particles also show a prominent peak at high eccentricities ($e >$ 0.85) as seen in Figure~\ref{fig:ELz_auriga} (middle panel).  Comparing the observed RRL eccentricities with GES-18, we estimate that $\sim$9 $\pm$ 2\% of the RRLs in the inner-central halo originated from GES.

The RRL signature of GES is therefore not as prominent in the inner-central halo as in the solar vicinity, but the excess of RRL on eccentric orbits, as well as the fraction of RRLs with integrals of motions consistent with GES, is consistent with belonging to GES debris in the inner Galaxy.
With the release of {\it Gaia} DR4, the number of RRLs with 3D velocities and 3D positions will be around $\sim$20,000, allowing a similar analysis to be carried out, but with a significantly larger sample and using a RRL sample that extends from the solar vicinity to the bulge and beyond.  
In the near future, the 4MOST spectrograph will increase this sample, as it delivers radial velocities and metallicities of as 100,000 RRLs over the southern sky through the 4MOST Gaia RR Lyrae Survey \citep[4GRoundS][]{ibata23}.  With spectroscopic abundances, it may be possible to separate the different accretion events that overlap in the inner Galaxy and allow a probe of the earliest accretion events that have today fallen deep into the potential well of the Milky Way. 


\begin{acknowledgments}
AMK acknowledges support from grant AST-2009836 and AST-2408324 from the National Science Foundation.  
AMK, KRC, JH, KD acknowledge the M.J. Murdock Charitable Trust's support through its RAISE (Research Across Institutions for Scientific Empowerment) program.  
Funded by the Deutsche Forschungsgemeinschaft (DFG, German Research Foundation) under Germany's Excellence Strategy -- EXC 2094-390783311.
EAT acknowledges financial support from ANID “Beca de Doctorado Nacional” 21220806.
AM acknowledges support from the ANID FONDECYT Regular grant 1251882, from the ANID BASAL project FB210003, and funding from the HORIZON-MSCA-2021-SE-01 Research and Innovation Programme under the Marie Sklodowska-Curie grant agreement number 101086388.
This research was supported by the Munich Institute for Astro-, Particle and BioPhysics (MIAPbP) which is funded by the Deutsche Forschungsgemeinschaft (DFG, German Research Foundation) under Germany's Excellence Strategy – EXC-2094 – 390783311.
We have used simulations from the Auriga Project public data release (https://wwwmpa.mpa-garching.mpg.de/auriga/data).
We thank Robert Grand for his help and comments regarding Au-18.

This work has made use of data from the European Space Agency (ESA) mission {\it Gaia} (\url{https://www.cosmos.esa.int/gaia}), processed by the {\it Gaia} Data Processing and Analysis Consortium (DPAC, \url{https://www.cosmos.esa.int/web/gaia/dpac/consortium}). Funding for the DPAC has been provided by national institutions, in particular the institutions participating in the {\it Gaia} Multilateral Agreement.

\end{acknowledgments}

{}

\end{document}